\documentclass[11pt]{article}

\begin{document}

\begin{center}
{\huge Spatial solitons in a medium composed of self-focusing and
self-defocusing layers}

\vspace{12mm}

{\large Javid Atai$^{1} $ and Boris A. Malomed$^{2} $}\\[0pt]
\vspace{3mm}

\emph{$^{(1)} $School of Electrical and Information Engineering, University
of Sydney, NSW 2006, Australia}\\[0pt]
$^{(2)} $\emph{Department of Interdisciplinary Studies, Faculty of
Engineering, Tel Aviv University, Tel Aviv 69978, Israel}\\[0pt]
\vspace{15mm}

\textbf{Abstract\vspace{5mm}}
\end{center}

We introduce a model combining Kerr nonlinearity with a periodically
changing sign ({}``nonlinearity management{}'') and a Bragg grating (BG).
The main result, obtained by means of systematic simulations, is presented
in the form of a soliton's stability diagram on the parameter plane of the
model; the diagram turns out to be a universal one, as it practically does
not depend on the soliton's power. Moreover, simulations of the nonlinear
Schr\"{o}dinger (NLS) model subjected to the same {}``nonlinearity
management{}'' demonstrate that the same diagram determines the stability of
the NLS solitons, unless they are very narrow. The stability region of very
narrow NLS solitons is much smaller, and soliton splitting is readily
observed in that case. The universal diagram shows that a minimum non-zero
average value of Kerr coefficient is necessary for the existence of stable
solitons. Interactions between identical solitons with an initial phase
difference between them are simulated too in the BG model, resulting in
generation of stable moving solitons. A strong spontaneous symmetry breaking
is observed in the case when in-phase solitons pass through each other due
to attraction between them. \newline

\newpage

\section{Introduction and formulation of the model}

Gap solitons are solitary waves in various nonlinear dispersive media, where
the spectrum of linear excitations contains a gap \cite{review}. Since
solitons of this kind had been observed \cite{experiment} in a short piece
of a nonlinear optical fiber equipped with a Bragg grating (BG), which gives
rise to the gap-bearing spectrum, they have attracted a great deal of
attention in nonlinear optics. A usual model of the BG fiber is based on a
system of two linearly coupled evolution equations \cite{AW,CJ}: 
\begin{eqnarray}
iu_{t}+iu_{x}+\left( \frac{1}{2}|u|^{2}+|v|^{2}\right) u+v & = & 0,
\label{uusual} \\
iv_{t}-iv_{x}+\left( \frac{1}{2}|v|^{2}+|u|^{2}\right) v+u & = & 0.
\label{vusual}
\end{eqnarray}
Here, $u(x,t) $ and $v(x,t) $ are local amplitudes of the counterpropagating
electromagnetic waves, the coefficient of the linear coupling, induced by
the resonant Bragg scattering, is normalized to be $1 $, and the cubic terms
account for the self-phase modulation (SPM) and cross-phase modulation (XPM)
effects induced by the Kerr nonlinearity.

The system of Eqs. (\ref{uusual}) and (\ref{vusual}) is frequently called a
generalized massive Thirring model. The Thirring model proper \cite{Thirring}
corresponds to the case when the SPM terms are dropped (only in that case,
the system is exactly integrable by means of the inverse scattering
transform).

Although the system (\ref{uusual}) and (\ref{vusual}) is not integrable, it
gives rise to a family of exact gap-soliton solutions \cite{AW,CJ}. It has
been demonstrated that a part of the family is stable, and another part is
unstable \cite{Rich,stability}. Exact gap-soliton solutions can also be
found, and their stability can be investigated, in generalized versions of
the model (\ref{uusual}), (\ref{vusual}), which describe a system of
parallel-coupled nonlinear and linear fibers with BG written on both of them 
\cite{dual1}, including also an especially interesting case when BG is
written only on the linear core, so that the nonlinearity and BG are
separated \cite{dual2}.

There is a principal difference between the gap solitons and the commonly
known solitons in ordinary single-mode nonlinear optical fibers (without
BG), described by the nonlinear Schr\"{o}dinger (NLS) equation \cite{Agrawal}
. In the usual NLS model, a soliton exists as a result of the balance
between the self-focusing SPM nonlinearity and anomalous temporal
dispersion. If the nonlinearity is self-defocusing, while the dispersion
remains anomalous, bright solitons do not exist. However, in the model of
the nonlinear optical fiber equipped with BG, the sign of the nonlinearity
does not matter, as the effective dispersion (or diffraction, see below)
induced by the grating includes both normal and anomalous branches, hence
either of them will be able to support solitons. The latter circumstance
suggests to consider a modified version of the standard model (\ref{uusual}),
(\ref{vusual}) where nonlinearity's sign may change.

The simplest possibility to realize the sign-changing nonlinearity is to
take it as a combination of cubic and quintic terms with opposite signs.
This modification of the standard model was considered in Ref. \cite{we}. As
well as in the case of the system (\ref{uusual}), (\ref{vusual}), stationary
soliton solutions of the modified system were found in an exact analytical
form, while their stability was studied by means of numerical simulations.
It has been concluded that the family of gap solitons in the modified system
is drastically different from that in its standard counterpart: the family
splits into two disjoint parts, each being dominated by one of the two
nonlinear terms of the opposite signs (in accordance with what might be
expected), and a part of each subfamily is stable.

A different possibility to study the effect of the sign-changing
nonlinearity on the gap solitons is to introduce a model with the
nonlinearity being represented by the cubic term only, whose sign is
changing periodically as a function of the evolution variable (in analogy to
dispersion management, see, e.g., Refs. \cite{first,DM}, this may be called
{}``nonlinearity management{}''). In the temporal domain {[}i.e., in a
BG-carrying optical fiber, see Eqs. (\ref{uusual}) and (\ref{vusual}){]},
this implies that the nonlinearity must periodically change its sign in
time, which is not a physically realistic assumption. However, a similar
arrangement can easily be implemented in the \textit{spatial domain}, i.e.,
for stationary light beams propagating across a layered structure in a
planar nonlinear waveguide.

It is worth mentioning that stable three-wave spatial gap solitons in a
model of a planar waveguide, in which BG was realized as a system of
parallel scores, were found in Ref. \cite{Mak}. However, in that work the
nonlinearity was quadratic {[}also known as $\chi ^{(2)} $ type{]}, and it
was not subject to any modulation. On the other hand, stable
(2+1)-dimensional (cylindrical) spatial solitons in a model of a bulk
(three-dimensional) layered medium with the periodically varying sign of the
cubic nonlinearity, but without BG, have recently been reported in Ref. \cite
{Isaac} (earlier, it was demonstrated in Ref. \cite{layered} that a layered
structure in the bulk medium helps to stabilize the cylindrical solitons
even in the case when the size of the nonlinearity coefficient varies
between the layers, without changing its sign). We do not aim to discuss in
detail here how the layers with the opposite signs of the Kerr coefficient
can be realized. However, it is relevant to point out that one possibility
is to induce an effective Kerr coefficient via the cascading mechanism,
based on the $\chi ^{(2)} $ nonlinearity, where the sign of the effective $
\chi ^{(3)} $ coefficient can be controlled by means of the $\chi ^{(2)} $
mismatch parameter \cite{Frank}.

Thus, the model to be considered has the form {[}cf. Eqs. (\ref{uusual}) and
(\ref{vusual}){]}
\begin{eqnarray}
iu_{z}+iu_{x}+\gamma (z)\left( \frac{1}{2}|u|^{2}+|v|^{2}\right) u+v & = & 0,
\label{u} \\
iv_{z}-iv_{x}+\gamma (z)\left( \frac{1}{2}|v|^{2}+|u|^{2}\right) v+u & = & 0,
\label{v}
\end{eqnarray}
where $z $ is the propagation distance, which plays the role of the
evolution variable instead of time in Eqs. (\ref{uusual}) and (\ref{vusual}
), and $x $ is the transverse coordinate in the layered planar waveguide.
The form of Eqs. (\ref{u}) and (\ref{v}) implies that the carrier wave
vectors of the two waves, which are resonantly reflected into each other by
BG form equal angles with the $z $ axis. The reflecting scores which form
the BG on the planar waveguide are oriented normally to the $z $ axis. The
usual diffraction in the waveguide is neglected {[}similar to the
negligibility of the intrinsic group-velocity dispersion of the optical
fiber in the temporal-domain model based on Eqs. (\ref{uusual}) and (\ref
{vusual}){]}, as it is assumed that BG gives rise to a much stronger
diffraction, cf. a similar situation in the work \cite{Mak}.

Although Eqs. (\ref{u}) and (\ref{v}) are $z $-dependent, they conserve the
net power, 
\begin{equation}  \label{E}
P\equiv \int _{-\infty }^{+\infty }\left( |u(x)|^{2}dx+|v(x)|^{2}\right) dx.
\end{equation}
The equations can be represented in a Hamiltonian form, but, unlike the net
power, the corresponding Hamiltonian is not a dynamical invariant, due to
the presence of the explicit $z $-dependence in it.

The layered structure of the waveguide assumes that the Kerr coefficient $
\gamma (z) $ takes positive and negative values $\gamma _{+} $ and $\gamma
_{-} $ in the alternating layers (cf. Ref. \cite{Isaac}): 
\begin{equation}  \label{gamma}
\gamma (z)=\left\{ 
\begin{array}{cc}
\gamma _{+}, & \mathrm{if}\, \, \, 0<z<L_{+} \\ 
\gamma _{-}, & \mathrm{if}\, \, \, L_{+}<z<L_{+}+L_{-},
\end{array}
\right.
\end{equation}
which is repeated periodically with the period $L\equiv L_{+}+L_{-} $. Using
the scaling invariance of Eqs. (\ref{u}) and (\ref{v}), one may always
impose the following normalization conditions: 
\begin{equation}  \label{normalization}
L_{+}+L_{-}\equiv 1,\, L_{+}\gamma _{+}+L_{-}|\gamma _{-}|\equiv 1.
\end{equation}
Thus, the model contains two irreducible control parameters, which may be
selected as, e.g., $L_{+} $ and $\gamma _{+} $, while the other parameters
can be found from Eqs. (\ref{normalization}), 
\begin{equation}  \label{minus}
L_{-}=1-L_{+},\, \gamma _{-}=-\left( 1-L_{+}\gamma _{+}\right) /\left(
1-L_{+}\right) .
\end{equation}
Note that corresponding average value of the Kerr coefficient is 
\begin{equation}  \label{average}
\overline{\gamma }\equiv \frac{L_{+}\gamma _{+}+L_{-}\gamma _{-}}{L_{+}+L_{-}
}=2L_{+}\gamma _{+}-1.
\end{equation}
As we are interested in the sign-changing model, with $\gamma _{-}<0 $, we
will confine the consideration to the case $L_{+}\gamma _{+}\leq 1 $, which
is equivalent to $\gamma _{-}\leq 0 $ according to Eqs. (\ref{minus}).

As is known, broad-gap-soliton solutions to Eqs. (\ref{uusual}) and (\ref
{vusual}) are asymptotically equivalent to broad NLS solitons. This suggests
to consider, parallel to the model (\ref{u}), (\ref{v}), also the NLS
equation with the nonlinearity coefficient subjected to the same periodic
modulation as in Eq. (\ref{gamma}), 
\begin{equation}  \label{NLS}
iu_{z}+\frac{1}{2}u_{xx}+\gamma (z)\left| u\right| ^{2}u=0,
\end{equation}
where the variable $x $ may be realized as either the transverse coordinate
in the planar waveguide, or the standard temporal coordinate in an optical
fiber \cite{Agrawal}, $z $ being the propagation distance in either case.
Comparing the results for the gap and NLS solitons in the two models will be
quite helpful in realizing the generality of the conclusions presented below.

The rest of the paper is organized as follows. In section 2 we present the
main results obtained in this work, in the form of a stability diagram for
the solitons in the BG model, its remarkable features being that the
stability region virtually does not depend on the soliton's power, and a
non-zero average value of the Kerr coefficient is necessary for the
existence of stable solitons. Similar results for the NLS model are
presented, in a brief form (as they differ from results for the gap solitons
only in the case when the solitons are very narrow), in section 3. In
section 4, results produced by interactions between two stable gap solitons
with a phase difference between them are presented; in particular, stable
moving solitons are readily generated by the interaction. The paper is
concluded by section 5.

\section{The stability diagram for the gap solitons}

The main issue of this work is the existence of stable solitons in the
model. Obviously, a soliton may only be a pulsating one ({}``breather{}''),
due to the periodic inhomogeneity of the system. For the same reason, the
problem is too difficult for an analytical treatment. In principle, one may
try to develop an analytical approach for pulsating solitons based on the
variational approximation, as it was done successfully in the case of the
layered medium described by the (2+1)-dimensional NLS equation with the
periodically changing sign of the Kerr coefficient \cite{Isaac}; however,
unlike the NLS solitons, the variational approximation for the gap solitons
is very complex even without the periodic modulation, see Ref. \cite{Rich}.
Therefore, we rely on direct numerical simulations of Eqs. (\ref{u}) and (
\ref{v}), with the coefficient $\gamma (z) $ defined as per Eqs. (\ref{gamma}
) and (\ref{minus}).

At $z=0 $, we used the well-known exact gap-soliton solution of the standard
model (\ref{uusual}) and (\ref{vusual})\cite{AW,CJ}: 
\begin{equation}  \label{solBragg}
\begin{array}{rl}
u_{0}(x)= & \left( 1/\sqrt{3}\right) \left( \sin \theta \right) \mathrm{sech}
\left( x\sin \theta -\frac{1}{2}i\theta \right) , \\ 
v_{0}(x)= & -\left( 1/\sqrt{3}\right) \left( \sin \theta \right) \mathrm{sech
}\left( x\sin \theta +\frac{1}{2}i\theta \right) ,
\end{array}
\end{equation}
where $\theta $ is an internal parameter of the soliton family, which takes
values $0< $ $\theta <\pi $. As is known, these solitons are stable in the
interval $0< $ $\theta < $ $\theta _{\mathrm{cr}} $, where the border
(critical) value is slightly larger than $\pi /2 $, $\theta _{\mathrm{cr}
}\approx 0.506\cdot \pi $ \cite{Rich,stability}). We ran the simulations for
a fixed value of the parameter $\theta $ in the initial configuration (\ref
{solBragg}), while the model's control parameters $\gamma _{+} $ and $L_{+} $
were gradually varied, subject to the above-mentioned constraint $
L_{+}\gamma _{+}\leq 1 $. Then, the same was done for other values of $
\theta $.

By means of these simulations we have been able to identify a stability
region for the solitons in the parameter plane $\left( L_{+},\gamma
_{+}\right) $, which is shown in Fig. 1. The upper boundary of the stability
region shown in Fig. 1 is $L_{+}\gamma _{+}=1 $, which, as was explained
above, limits the case considered in this work, as the local Kerr
coefficient does not change sign above this boundary . The upper boundary
itself corresponds, as per Eqs. (\ref{minus}), to a system in which
nonlinear layers of the width $L_{+} $ alternate with linear ones (having $
\gamma _{-}=0 $) of the width $L_{-} $. The simulations demonstrate that
everywhere on this boundary, the solitons are stable, and they remain stable
above the boundary, so that it only bounds the region that we aim to
consider in this work. It is noteworthy that the system composed of
alternating nonlinear and linear segments is similar to the recently
introduced model of a \textit{split-step} fiber-optic link, which also
readily gives rise to stable solitons, despite its strongly heterogeneous
structure \cite{Radik}.

The lower boundary of the stability region in Fig. 1 is rather close to a
hyperbola, the product $L_{+}\gamma _{+} $ along this boundary taking values
between $0.65 $ and $0.70 $. For comparison, the dashed curve in Fig. 1
shows the hyperbola $L_{+}\gamma _{+}=1/2 $, along which the average Kerr
coefficient (\ref{average}) exactly vanishes. The finite separation between
the lower stability boundary and the dashed curve can be measured by the
average value $\overline{\gamma } $ of the Kerr coefficient (\ref{average}).
The smallest value for $\overline{\gamma } $ found on the lower boundary is
approximately $0.3 $. Thus, stable solitons are not possible in a system
where the average value of Kerr coefficient is zero. This is, incidentally,
a noteworthy difference from the dispersion-managed fiber-optic systems,
where stable solitons have been found in the case when the average
dispersion exactly vanishes \cite{DM}.

The left vertical boundary of the stability region in Fig. 1 at $L_{+}=0.2 $
is not a real stability border. It only bounds a range in which the
systematic simulations were performed, which is $0.2\leq L_{+}<1.0 $.

The result described above is similar to that obtained in Ref. \cite{Isaac}
for the (2+1)-dimensional solitons in a layered medium without BG: there
too, a \emph{finite positive} average value of the Kerr coefficient was
necessary for the existence of any soliton, stable or unstable. It will be
shown in the next section that the same result is also true for the
one-dimensional layered NLS system. On the other hand, a difference of the
present model from the NLS one is that the sign of the nonlinearity is not
crucial (as it was explained in detail above). Indeed, the substitution $
u^{\ast }\equiv \widetilde{u} $, $v^{\ast }=-\widetilde{v} $ transforms Eqs.
(\ref{u}) and (\ref{v}) into themselves, but with the opposite sign of the
Kerr coefficient $\gamma (z) $, i.e., with 
\begin{equation}  \label{-gamma}
\gamma (z)=\left\{ 
\begin{array}{cc}
\widetilde{\gamma }_{+}\equiv \left| \gamma _{-}\right| , & \mathrm{if}\, \,
\, 0<z<\widetilde{L}_{+}\equiv L_{-}, \\ 
\widetilde{\gamma }_{-}\equiv -\gamma _{+}, & \mathrm{if}\, \, \, \widetilde{
L}_{+}<z<L_{+}+L_{-},
\end{array}
\right. .
\end{equation}
Therefore, there must exist another stability area in the region where the
average Kerr coefficient is negative, i.e., $L_{+}\gamma _{+}<1/2 $.
However, the distribution ({}``nonlinearity-management map{}'') of the Kerr
coefficient corresponding to Eq. (\ref{-gamma}) does not satisfy the second
normalization condition (\ref{normalization}). Therefore, it is necessary to
additionally redefine the wave fields, so that 
\begin{equation}  \label{tilde}
\left( \widetilde{u},\widetilde{v}\right) \equiv \frac{\left( \widetilde{
\widetilde{u}},\widetilde{\widetilde{v}}\right) }{\sqrt{\widetilde{L}_{+}
\widetilde{\gamma }_{+}+\widetilde{L}_{-}|\widetilde{\gamma }_{-}|}}=\left( 
\widetilde{\widetilde{u}},\widetilde{\widetilde{v}}\right) \sqrt{\frac{
1-L_{+}}{L_{+}+\gamma _{+}-2L_{+}\gamma _{+}}}\, ,
\end{equation}
where $\widetilde{L}_{\mp }\equiv L_{\pm } $, and Eqs. (\ref{minus}) and (
\ref{-gamma}) were taken into regard. Finally, with regard to the
transformation (\ref{tilde}), the second stability region (not shown in Fig.
1), corresponding to the negative average value of the Kerr coefficient, is
obtained from the region shown in Fig. 1 by the transformation 
\[
\left( L_{+};\gamma _{+}\right) \rightarrow \left( 1-L_{+};\frac{
1-L_{+}\gamma _{+}}{L_{+}+\gamma _{+}-2L_{+}\gamma _{+}}\right) .
\]

In the cases when the initial pulse (\ref{solBragg}) gives rise to a stable
soliton, the formation of the soliton is accompanied by emission of
radiation, and, possibly, by generation of a small additional pulse. An
example of stable soliton formation is shown in Fig. 2. The emission of
radiation is conspicuous in the case when a stable soliton is in the
proximity of the lower stability border. On the other hand, in the unstable
case the initial pulse completely decays into radiation; a typical example
is shown in Fig. 3. Strong symmetry breaking evident in Figs. 2 and 3 has,
plausibly, the same reason as in Fig. 7(a), which is discussed below.

The stability diagram displayed in Fig. 1 was obtained by simulations of
Eqs. (\ref{u}) and (\ref{v}) with the initial conditions (\ref{solBragg})
having $\theta =0\allowbreak .\, \allowbreak 484\cdot \pi $, which is close
to the above-mentioned stability-limiting value $\theta _{\mathrm{cr}
}\approx 0.506\cdot \pi $ of the standard model (\ref{uusual}), (\ref{vusual}
) which has $\gamma \equiv 1 $. The stability region may, of course, depend
on the initial value of $\theta $, or, in other words, on the initial value
of the soliton's power, defined by the expression (\ref{E}), which is $P_{
\mathrm{sol}}=\left( 4/3\right) \theta $ for the configuration (\ref
{solBragg}). As for the dependence of the stability region {[}defined in
terms of the parameter plane $\left( L_{+},\gamma _{+}\right) $, as in Fig. 1
{]} on soliton's power, one may naturally expect that the dependence is
strongest when the initial value of $\theta $ is close to the
above-mentioned value $\theta _{\mathrm{cr}} $, which is the stability
border for solitons in the ordinary BG model (\ref{uusual}), (\ref{vusual}).
Our simulations show that, in fact, the stability region shows virtually no
sensitivity to the variation of $\theta $: for example, decreasing $\theta $
from the value $\theta =\cos ^{-1}(0.05)\approx 0\allowbreak .\, \allowbreak
484\cdot \pi $, corresponding to Fig. 1, to $\theta =\cos ^{-1}\left(
0.15\right) \approx 0.452\cdot \pi $, produces no visible change whatsoever
in the shape of the stability region. Because the region shows no
sensitivity to the initial power in the case when the sensitivity might be
strongest, we conjecture (which is also corroborated by additional
simulations for essentially smaller values of $\theta $) that the stability
diagram shown in Fig. 1 is a universal one, without any tangible dependence
upon the soliton's power.

In this connection, it may be relevant to mention a somewhat similar
stability property of the gap solitons in the ordinary model (\ref{uusual})
and (\ref{vusual}): besides the zero-velocity ones given by Eqs. (\ref
{solBragg}), more general analytical solutions for solitons moving at an
arbitrary velocity $V $, which takes values in the interval $-1<-V<1 $, are
also known \cite{AW,CJ}. This generalization of the exact soliton solutions
is not trivial, as Eqs. (\ref{uusual}) and (\ref{vusual}) do not feature any
invariance with respect to the change of the reference frame; for the same
reason, the stability limit for the moving solitons cannot be obtained by
any straightforward transformation from that for the zero-velocity ones.
Nevertheless, direct numerical results \cite{stability} show that, although
the corresponding limit value $\theta _{\mathrm{cr}} $ does depend on $V $,
the dependence is very weak: as $V $ increases from $0 $ to $1 $, the
stability boundary gradually moves from the above-mentioned value $\theta _{
\mathrm{cr}}(V=0)\approx 0.506\cdot \pi $ to $\theta _{\mathrm{cr}
}(V=1)\approx 0.508\cdot \pi $.

\section{Stability of solitons in the NLS equation with spatially modulated
nonlinearity}

It is well known that the BG model based on Eqs. (\ref{uusual}) and (\ref
{vusual}) is related to the NLS equation: in the limit $\theta \rightarrow 0 
$, a small-amplitude broad gap soliton given by Eqs. (\ref{solBragg}) is
asymptotically equivalent to the NLS soliton. Therefore, in the limit of
small $\theta $ Eqs. (\ref{u}) and (\ref{v}) are tantamount to the NLS
equation (\ref{NLS}) with the periodically modulated nonlinear coefficient.
On the other hand, for solitons with a small amplitude $\theta $ and a large
width $\sim 1/\theta $, a characteristic diffraction length, which
determines the dynamical scale for the evolution of the soliton, is $\sim
1/\theta ^{2} $. This length is much larger than the modulation period of
the Kerr coefficient, which is $L\equiv 1 $ in the present notation, hence
the small-amplitude broad soliton will feel only the average value (\ref
{average}) of the Kerr coefficient. This, in turn, means that the broad
solitons should be stable in both the BG and NLS models for any positive
value of the average Kerr coefficient $\overline{\gamma } $.

The latter conclusion appears to be in contradiction with the results
presented in Fig. 1, where, as was stressed above, for all the stable
solitons the average Kerr coefficient $\overline{\gamma } $ was found to
exceed the minimum value $\approx 0.3 $. However, direct simulations of very
broad solitons in either model (the BG or NLS one) are quite difficult to
run, and they do not readily produce definite results, as very long
propagation is necessary, in view of the large value of the above-mentioned
diffraction length for these solitons. Besides that, it is not easy to
distinguish a small-amplitude soliton from radiation waves which it sheds
off in the course of the evolution. Lastly, this issue, i.e., finding exact
stability limits for very broad gap and NLS solitons, is of no practical
interest, as experiments with such solitons would require very large samples
of the layered waveguide and would therefore be extremely difficult, not
promising anything for applications. Therefore, we did not try to
investigate this issue in detail.

Nevertheless, for the sake of the comparison between the two models just in
the case when they are really different, it is of interest to generate a
stability diagram for (not-very-broad) NLS solitons, of the same type as the
diagram displayed in Fig. 1. To this end, systematic simulations of Eq. (\ref
{NLS}) were performed, with $\gamma (z) $ taken as per Eqs. (\ref{gamma}), (
\ref{minus}). The simulations began with the initial condition 
\begin{equation}  \label{initial}
u_{0}(x)=\eta \, \mathrm{sech}\left( \eta x\right) ,
\end{equation}
that would generate an exact soliton in the NLS equation with $\gamma \equiv
1 $. In view of the above-mentioned difficulties with simulations of very
broad solitons, definite results were obtained for the initial
configurations (\ref{initial}) with $\eta \geq \eta _{\min }=0.5 $ {[}for $
\eta =0.4 $, the situation is very similar, but residual evolution of the
soliton does not completely cease during the simulation time; note the $\eta
=0.5 $ corresponds, roughly, to $\theta =\pi /6 $ in the case of the gap
soliton (\ref{solBragg}), if the comparison is made by the soliton's width{]}
.

The result is that, for all the moderately narrow initial solitons of the
type (\ref{initial}), the thus obtained stability diagram is virtually \emph{
indistinguishable} from its counterpart in Fig. 1. This finding once again
stresses the universal character of the stability diagram presented in Fig.
1, as it applies to both models in the case when they are really different.

The difference between the BG and NLS models becomes drastic if the
amplitude $\eta $ of the initial pulse (\ref{initial}) is high (recall that
the amplitude of the NLS soliton may be indefinitely large, while the
amplitude of the gap solitons (\ref{solBragg}), which is $\left( 2/\sqrt{3}
\right) \sin \left( \theta /2\right) $, does not exceed the maximum value \ $
2/\sqrt{3} $ ). It is obvious that, for very narrow solitons with large $
\eta $, whose diffraction length $\sim 1/\eta ^{2} $ is much smaller than
the modulation period $L=1 $, the periodic change of the nonlinearity sign,
as per Eq. (\ref{gamma}), is a strong perturbation that may destroy the
soliton. Indeed, running the simulations for the NLS equation (\ref{NLS})
with the initial condition (\ref{NLS}), we have observed that the stability
region strongly shrinks (compared with that shown in Fig. 1) in the case of
large $\eta $. Moreover, even if the evolution of the initial pulse (\ref
{initial}) with large $\eta $ results in the appearance of stable solitons,
they were frequently produced by \emph{splitting} of the initial pulse. The
situation was classified as a stable one if stable solitons finally appeared
in any fashion.

An example of the thus defined stability diagram for the NLS model (\ref{NLS}
) with a large initial amplitude $\eta =5 $ is displayed in Fig. 4, and an
example of the splitting of the initial pulse into two secondary solitons
with a much smaller amplitude is presented in Fig. 5. In the case $\eta =5 $
, the splitting was always observed if, for instance, the parameter $L_{+} $
was fixed to be $0.4 $. We did not aim to find the smallest value of the
initial amplitude which gives rise to the splitting, as this would require
running extremely long simulations (it is plausible that the splitting
distance diverges around that smallest value of $\eta $).

In some other cases, a stable soliton, with an amplitude essentially smaller
than the large initial amplitude, appears as a result of what may be
interpreted as splitting of the initial pulse into a central soliton and two
very weak additional solitons which move aside. A typical example of this
outcome is shown in Fig. 6.

\section{Interactions between solitons and generation of moving solitons}

Coming back to gap solitons, we note that, as the standard model (\ref
{uusual}), (\ref{vusual}) is not integrable, interactions between the gap
solitons are not elastic. This has already been demonstrated in the first
direct simulations of the model \cite{AW}. One may, therefore, expect that
interactions between stable gap solitons in Eqs. (\ref{u}) and (\ref{v})
will not be elastic either.

We studied interactions between gap solitons, taking two identical ones,
giving them an initial phase difference $\Delta \phi $, and placing them at
some distance $\Delta x $ from each other. As a result, we observed that,
quite naturally, the in-phase solitons with $\Delta \phi =0 $ attract each
other, while out-of-phase ones with $\Delta \phi =\pi $ mutually repel, see
Figs. 7(a) and 7(c). In the intermediate case $\Delta \phi =\pi /2 $, the
solitons also repel each other, see Fig. 7(b).

Fig. 7 demonstrates two other important features. First, it clearly shows
that stable moving solitons also exists in the present model, and they can
be easily generated. Second, in the case when the two solitons initially
attract each other, and hence temporarily merge into a {}``lump{}'' {[}see
Fig. 7(a){]}, conspicuous spontaneous symmetry breaking is observed, and the
outcome of the interaction is inelastic: generation of an extra moving
(quasi-)soliton, along with some radiation, is clearly seen. In fact, this
spontaneous symmetry breaking, in the case of the initially attractive
interaction, is similar to an effect recently reported in Ref. \cite{dual2}.
It seems plausible that an explanation which was proposed in that paper
applies to the present case as well: the above-mentioned {}``lump{}'' is
subject to modulational instability \cite{Agrawal}, hence the amplification
of small random numerical perturbations by the instability gives rise to the
symmetry breaking.

Collisions between moving solitons may also lead to inelastic interactions,
but detailed simulations of such collisions are beyond the scope of this
paper.

\section{Conclusion}

In this work, we have introduced a model that describes a planar nonlinear
waveguide, equipped with a Bragg grating and composed of periodically
alternating self-focusing and self-defocusing layers. The main result,
obtained by means of systematic simulations of the evolution of solitons in
the model, was presented as the soliton's stability diagram on the parameter
plane of the model; it has been found that the diagram practically does not
depend on the soliton's power. An important feature revealed by the
stability diagram is that a minimum non-zero value of the average Kerr
coefficient is necessary for the existence of stable solitons.

To compare the result with the nonlinear Schr\"{o}dinger model, where the
Kerr coefficient is subjected to the same sign-changing modulation as in the
Bragg-grating model, the NLS model was also simulated in a systematic way.
It has been found that, for moderately narrow NLS solitons, the stability
diagram is virtually identical to that for the gap solitons, which stresses
the universal character of this diagram. For very narrow NLS solitons, the
stability region is much smaller, and, in the latter case, stable solitons
frequently appear as a result of splitting of the initial pulse.

Interactions between two identical stable gap solitons with an initial phase
difference between them were simulated too. As a result, stable moving
solitons were easily generated, and a strong spontaneous symmetry breaking
was observed in the case when two in-phase solitons pass through each other.

\newpage

\begin{center}
\textbf{\large Figure Captions}
\end{center}

Fig. 1. The stability diagram for solitons in the model based on Eqs. (\ref
{u}), (\ref{v}) and (\ref{gamma}), (\ref{minus}). The stability region is
bounded by the lower solid curve, while the upper curve, the hyperbola $
L_{+}\gamma _{+}=1 $, is a border of the parametric area where the local
Kerr coefficient periodically changes its sign. The dashed curve is the
hyperbola $L_{+}\gamma _{+}=1/2 $ along which the average value of
nonlinearity is zero.

Fig. 2. An example of the formation of a stable soliton when $L_{+}=0.5$ and 
$\gamma _{+}=1.7$. Only the $u$ component is shown.

Fig. 3. An example of the decay of the initial pulse (\ref{solBragg}) into
radiation at a point belonging to the unstable region in Fig. 1, with $
L_{+}=0.5 $ and $\gamma _{+}=1.2 $. Only the u component is shown.

Fig. 4. A narrow stability region in the nonlinear Schr\"{o}dinger model (
\ref{NLS}) where the Kerr coefficient periodically changes its sign as per
Eqs. (\ref{gamma}) and (\ref{minus}) for the case when the initial pulse is
given by Eq. (\ref{initial}) with $\eta =5 $.

Fig. 5. An example of splitting of the narrow pulse in the NLS model (\ref
{NLS}), (\ref{initial}) into two secondary solitons with a smaller amplitude
in the case $\eta =5 $. The parameters of the model are $L_{+}=0.35 $ and $
\gamma _{+}=2.74 $. The propagation distance shown in this figure is $200 $
(i.e., $200 $ modulation periods).

Fig. 6. An example of spitting of the narrow pulse in the NLS model (\ref
{NLS}), (\ref{initial}) into a central soliton with a smaller amplitude and
two very weak side pulses in the case $\eta =5 $. The parameters of the
model are almost the same as in Fig. 5, with a difference that $\gamma
_{+}=2.76 $.

Fig. 7. The interaction of two identical stable solitons, with the initial
phase difference $\Delta \phi $ and separation $\Delta x $ between their
centers, when $L_{+}=0.5 $, $\gamma _{+}=1.94 $. The solitons have been
generated from the initial configuration (\ref{solBragg}) with $\theta
\approx 0\allowbreak .\, \allowbreak 484\cdot \pi $ (the same value as the
one used in Fig. 1). The three cases displayed in the figure correspond to
(a) $\Delta \phi =0 $, $\Delta x=12 $; (b) $\Delta \phi =\pi /2 $, $\Delta
x=8 $; (c) $\Delta \phi =\pi $, $\Delta x=12 $. Only the u component is
shown.

\end{document}